\documentclass[showpacs,aps,pra,amsmath,superscriptaddress,twocolumn]{revtex4-1}

\usepackage{times}
\usepackage{graphicx}
\usepackage{amsmath}
\usepackage{amsfonts}
\usepackage{amssymb}
\usepackage{braket}
\usepackage{bm}
\usepackage{multirow}
\usepackage{color}
\usepackage[normalem]{ulem}
\usepackage{mathrsfs}
\usepackage{mathtools}
\usepackage{float}

\makeatletter

\newcommand{\Rmnum}[1]{\expandafter\@slowromancap\romannumeral #1@}
\makeatother

\usepackage[breaklinks]{hyperref}
\hypersetup{colorlinks=true, linkcolor=blue, citecolor=blue, filecolor=blue, urlcolor=blue}

\begin{document}

\title{Ultra-slow dynamics in a translationally invariant spin model
  for multiplication and factorization}

\author{Lei Zhang}

\affiliation{Physics Department, Boston University, Boston,
  Massachusetts 02215, USA}
  
\author{Stefanos Kourtis}

\affiliation{Physics Department, Boston University, Boston,
  Massachusetts 02215, USA}

\author{Claudio Chamon}

\affiliation{Physics Department, Boston University, Boston,
  Massachusetts 02215, USA}

\author{Eduardo R. Mucciolo}

\affiliation{Department of Physics, University of Central Florida,
  Orlando, Florida 32816, USA}

\author{Andrei E. Ruckenstein}

\affiliation{Physics Department, Boston University, Boston,
  Massachusetts 02215, USA}

\date{\today}

\begin{abstract}
We construct a model of short-range interacting Ising spins on a
translationally invariant two-dimensional lattice that mimics a
reversible circuit that multiplies or factorizes integers, depending
on the choice of boundary conditions. We prove that, for open boundary
conditions, the model exhibits no finite-temperature phase
transition. Yet we find that it displays glassy dynamics with
astronomically slow relaxation times, numerically consistent with a
double exponential dependence on the inverse temperature. The slowness
of the dynamics arises due to errors that occur during thermal
annealing that cost little energy but flip an extensive number of
spins. We argue that the energy barrier that needs to be overcome in
order to heal such defects scales linearly with the correlation
length, which diverges exponentially with inverse temperature, thus
yielding the double exponential behavior of the relaxation time.
\end{abstract}

\maketitle

\section{Introduction}
\label{sec:intro}

The nature of the slow relaxation of glassy systems in the absence of
disorder remains one of the most intriguing problems in condensed
matter physics. There are two classes of theories of glasses: one that
argues for a {\it bona fide} thermodynamic transition into a glass
phase at finite temperature, such as in the Random First Order
Transition
theory~\cite{Kirkpatrick-Thirumalai1987,Kirkpatrick-Wolynes1987,Kirkpatrick-Thirumalai-Wolynes1987};
and the other that views glassy behavior as a purely dynamical
phenomenon, in the absence of any finite temperature thermodynamic
transition (see Ref.~\onlinecite{Ritort2002} for a review). A notable
example of this latter scenario is provided by plaquette spin
models~\cite{Lipowski1997,Newman-Moore1999,Garrahan2000}, which are
known to display super-Arrhenius equilibration rates that scale as an
exponential of a power of the inverse temperature. It is then natural
to ask if there is a principled way of designing models --- without
disorder, frustration, or phase transitions --- that display
ultra-slow dynamics.

One common expectation is that the above questions can be addressed by
appealing to the correspondence between some statistical mechanics
models with \emph{translationally invariant} (short-range)
interactions and hard computational problems, such as the tiling of
the plane~\cite{Parisi2000,Gottesman2009}. It may seem natural to
speculate that, when regarded as descriptions of physical systems in
contact with thermal reservoirs, these models would display long
glassy relaxation rates that reflect the complexity of the
computation. Below we produce a counterexample which suggests that the
complexity class of a computation alone does not necessarily determine
the low temperature behavior of relaxation rates into the ground state
of the model representing the computation.

In this paper, we argue for the possibility of novel anomalously slow
relaxation rates that scale as a \emph{double} exponential of the
inverse temperature for a disorder-free spin model, inspired by
statistical mechanics representations of reversible classical
computational problems~\cite{Chamon2017}. This work was motivated by
our attempts to use thermal annealing in finding the ground state of a
model that, for appropriate boundary conditions, encodes either
multiplication of two integers or factorization of semiprimes. Our
initial simulations showed that in both the factoring problem and even
in the ``easy'' case of multiplication, a problem in complexity class
P, thermal annealing leads to a remarkably long time-to-solution,
consistent with a double exponential scaling with the inverse
temperature. (We note that another unsuccessful attempt at using
thermal annealing of a different statistical mechanics representation
of the factoring problem was described in Ref.~\cite{Henelius2011}.)

The principal result of the current paper is that the
double-exponential behavior of the relaxation time survives for open
boundary conditions, in which case we prove rigorously that the model
representing the multiplication circuit displays no thermodynamic
phase transition. While this ultra-slow relaxation rate represents a
negative result in the context of solutions of computational problems,
it points to a remarkable example of a translationally invariant,
short-range interacting model that displays astronomically slow glassy
dynamics and no thermodynamic transitions at any finite
temperature. The origin of the extremely slow relaxation rate can be
traced to the propagation of errors initiated at a single defect,
\emph{i.e.}, a bit mismatch, created during the annealing
process. Such defects cost little energy but result in an extensive
region of errors that cannot be removed except through healing from
the boundaries. We speculate that the activation barrier for this
healing process scales linearly with the correlation length, $\xi \sim
\exp(1/T)$, the temperature dependence of which reflects the absence
of a finite temperature phase transition.

The rest of this paper is organized as follows: In
Sec.~\ref{sec:model}, we detail the construction of a translationally
invariant two-dimensional spin model with local interactions, starting
from a multiplication circuit. In Sec.~\ref{sec:thermodynamics}, we
compute the partition function of this model and prove the absence of
finite temperature thermodynamic phase transitions. In
Sec.~\ref{sec:annealing}, we introduce two other computational models
(which also display no finite temperature transitions) that help us
highlight the special features of our principal model described in
Sec.~\ref{sec:model}; and we compare the resulting relaxation rates
computed through thermal annealing.  The different processes required
in healing thermally induced defects, which are responsible for the
qualitatively different behaviors of the relaxation rates of the three
models, are discussed in Sec.~\ref{sec:defects}, where we also
speculate on the origin of the double-exponentially slow
relaxation. Our conclusions are presented in
Sec.~\ref{sec:conclusion}.

\section{Model}\label{sec:model}

\begin{figure}
  \includegraphics[width=3.5in]{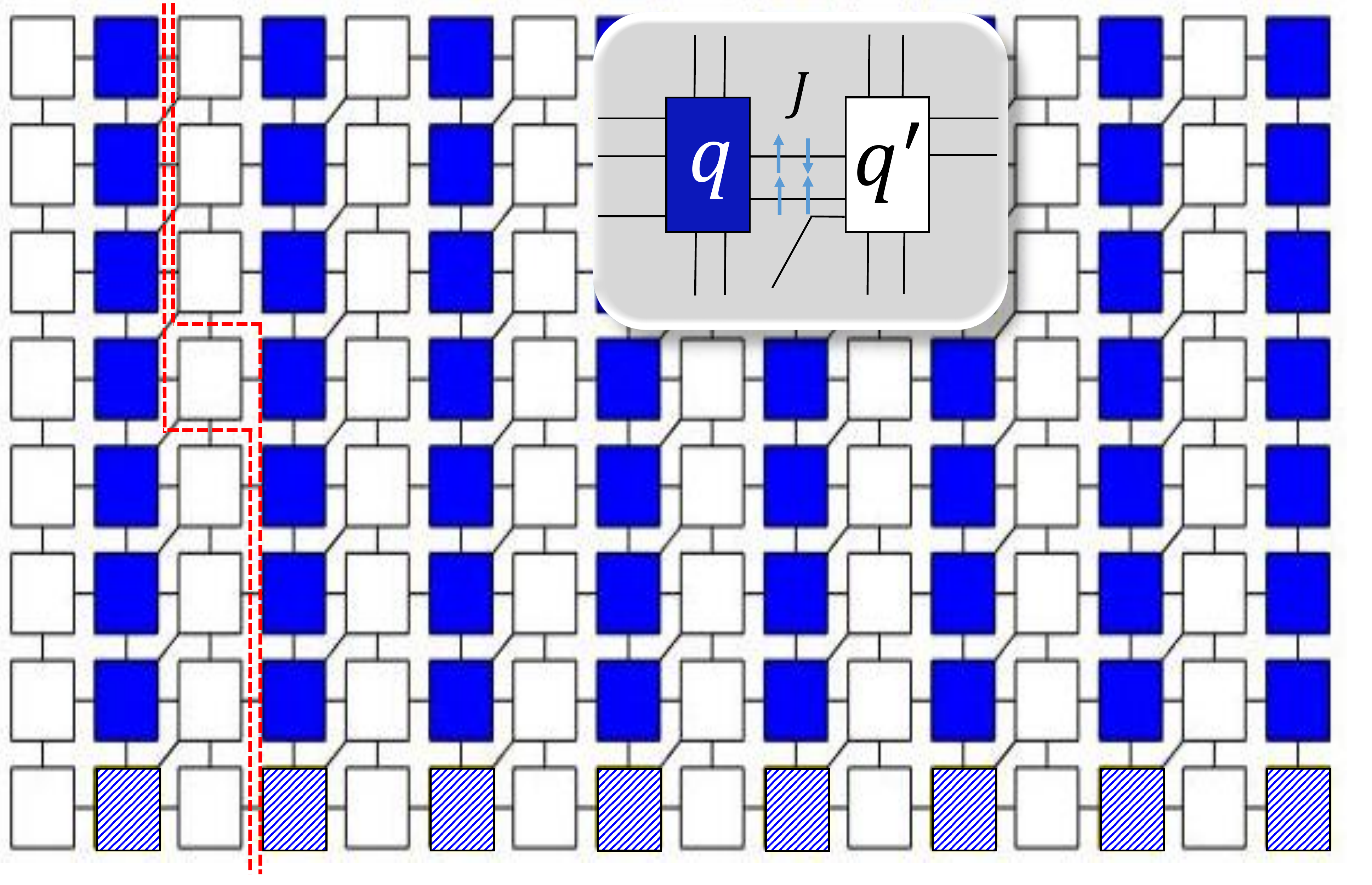}
\caption{Schematic of gates placed at the sites of a regular lattice
  with each gate connected to its neighbors. The inset depicts the
  details of the connection between neighboring gates: each link
  connects two spins (bits) representing one of the outputs of one
  gate and one of the inputs of its neighboring gate, respectively,
  The ferromagnetic interaction $J$ ensures that the output on one
  gate is compatible with the input to its neighbor. The $q$ and $q'$
  label the input/output configurations of the spins (bits) that
  satisfy the truth table for each of the two gates, respectively. Red
  dashed lines demonstrate the slicing discussed in
  Sec.~\ref{sec:thermodynamics}.
\label{fig:layers}}
\end{figure}

\begin{figure}
\includegraphics[width=3.5in]{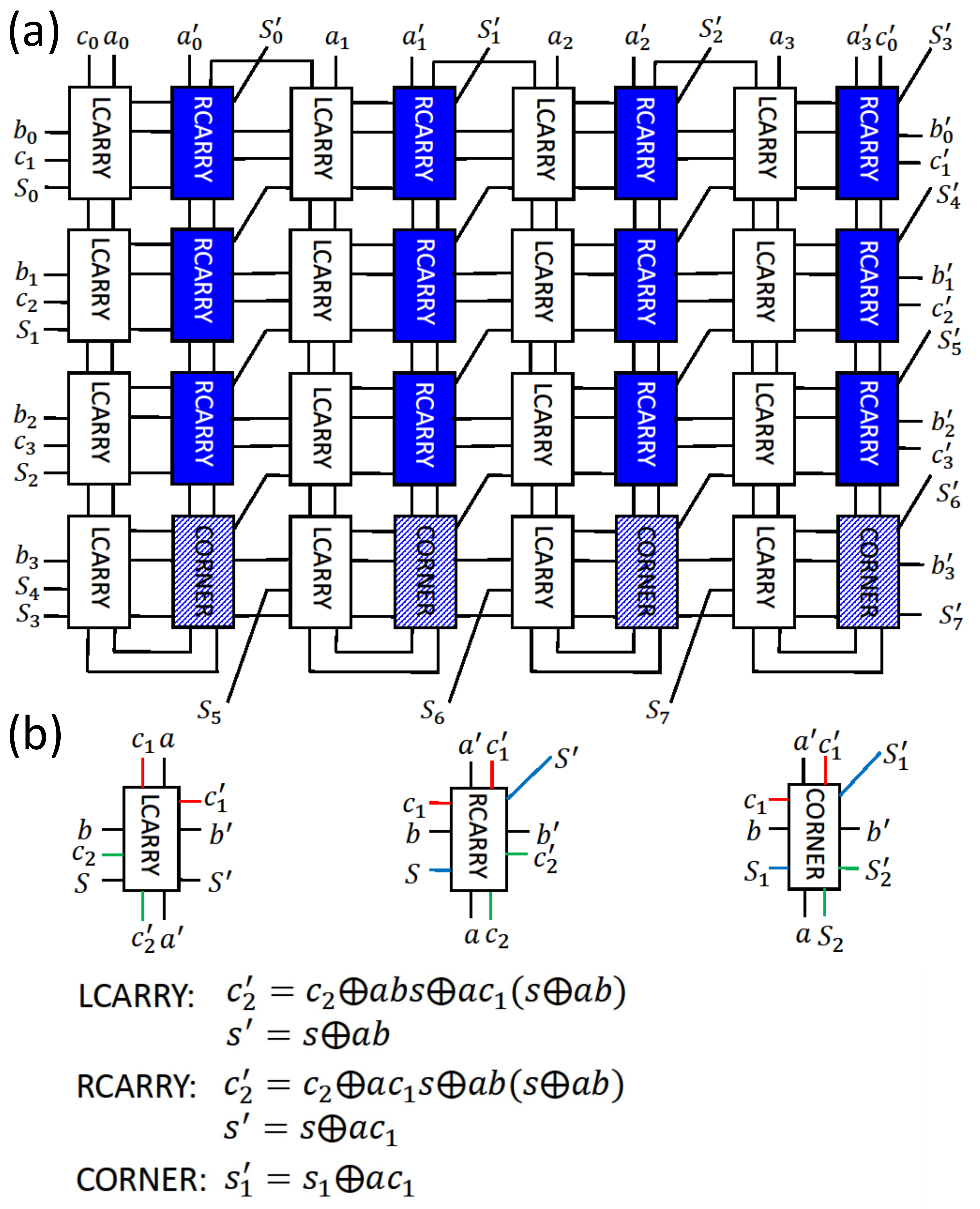}
\caption{(a) Schematic of the multiplication circuit for 4-bit numbers
  $a$ and $b$; (b) detailed structure of elementary gates
  $\mathsf{LCARRY}$, $\mathsf{RCARRY}$ and $\mathsf{CORNER}$ used in
  the circuit, along with their truth
  tables.}\label{fig:Circuit_2D_4bits}
\end{figure}

In this Section, we present a classical statistical mechanics spin
model on a two-dimensional lattice with short-range interactions that
encodes the functionality of a reversible multiplication circuit. The
two elements of the model are: (a) a regular lattice covered by logic
gates, the states of which are dictated by truth tables of allowed
configurations of input and output bits (or Ising spins) for each
gate; and (b) an interaction energy that penalizes neighboring gates
for which input/output states are incompatible with one another.

We illustrate the spin model in Fig.~\ref{fig:layers}, where the gates
(boxes) are connected so as to form a regular lattice. The inset
highlights two neighboring gates, with their connections to their
neighbors, where bit lines connect outputs to inputs. In the case
depicted (that is relevant to the multiplication circuit we deploy),
each gate has 10 line-connections, corresponding to 5 input and 5
output bits. The states $q$ and $q'$ of the gates take values
$0,1,\dots,2^5-1$, which label all possible 5-bit input states; here
we utilize reversible gates, so that the (5) output bits are uniquely
determined by the (5) input bits, and {\it vice-versa}. The
input/output relation (or truth table) for each gate is energetically
enforced, and we also add interactions that penalize configurations of
states where the inputs of one gate do not match the outputs of its
neighbors. This penalty enters in the form of a ferromagnetic
interaction, depicted in the inset of Fig.~\ref{fig:layers}, which
shows examples of parallel and anti-parallel spins corresponding,
respectively, to matched and unmatched bits between the two
neighboring gates.

In Fig.~\ref{fig:Circuit_2D_4bits} we show the specific circuit that
implements multiplication for 4-bit numbers. (Multipliers of $L$-bit
numbers are implemented in the same fashion.) This circuit is a
reversible logic version of the standard array multiplier. It utilizes
3 types of gates: $\mathsf{LCARRY}$, $\mathsf{RCARRY}$ and
$\mathsf{CORNER}$. Each of these gates has 5 inputs and 5 outputs,
with truth tables shown in panel (b) of
Fig.~\ref{fig:Circuit_2D_4bits}. It is at the boundaries of the system
where the inputs and outputs of the whole multiplication circuit are
written to or read from. The result of a multiplication of two $L$-bit
numbers, $a=2^{L-1}\,a_{L-1}+\dots+2\,a_1+a_0$ and
$b=2^{L-1}\,b_{L-1}+\dots+2\,b_1+b_0$, namely $S'=a\times b =
2^{2L-1}\,S'_{2L-1}+\dots+2\,S'_1+S'_0$, is placed in the $2L$-bit
registers, $S'_0, S'_1,...,S'_{(2L-1)}$.  The bits in the input
register $S$, as well as in the carrier input $c$, are all set to
0. (The output bits of the carrier $c'$ are also 0 at exit.)  Note
that, in this design, the total number of gates required for the
multiplication of two $L$-bit numbers is $N_{\mathrm{gates}}=2L^2$.

In order to construct the Hamiltonian of the spin model, which
formalizes the above discussion, we introduce the following
notation. We start by labeling the gates by an index $g$, and the
links or wires between gates by $\ell$. It is useful to define the
sets of input and output wires out of gate $g$, denoted by $w^{\rm
  in}(g)$ and $w^{\rm out}(g)$. Notice that if an output of gate $g$
is connected to an input of gate $g'$ by a wire $\ell$, then $\ell\in
w^{\rm out}(g)\cap w^{\rm in}(g')$. On each wire $\ell$ we define two
classical spins $\sigma^{\text{in}}_\ell$ and
$\sigma^{\text{out}}_\ell$, which we refer to as ``twin'' spins. These
spins, which represent the Boolean variables via the relation
$x^{\text{in,out}}_\ell = (1+\sigma^{\text{in,out}}_\ell)/2$, must
align ferromagnetically if the output of one gate is to match the
input of its neighbor. We shall use the shorthand notation
$\sigma^{\text{in}}(g)$ for the 5 variables $\sigma^{\text{in}}_\ell$
with $\ell \in w^{\rm in}(g)$, and $\sigma^{\text{out}}(g)$ for the 5
variables $\sigma^{\text{out}}_\ell$ with $\ell \in w^{\rm
  out}(g)$. Finally, we denote by $L_\partial$ the set of links $\ell$
at the boundary, which we use to fix inputs/outputs to the circuit
(these are wires at the boundary of the circuit).

With this notation, the Hamiltonian of the model is given by
\begin{subequations}
\begin{align}
  H
  =
  &{\ } H_{\mathrm{gates}} + H_{\mathrm{links}} + H_{\mathrm{boundary}}
  \,,\\
  H_{\mathrm{gates}}
  =
  &{\ } \sum_{g=1}^{N_{\mathrm{gates}}} \Delta_g\,{\overline T}_g[\sigma^{\text{in}}(g), \sigma^{\text{out}}(g)]
  \,,\\
  H_{\mathrm{links}}
  =
  &{\ } - \sum_{\ell} J_\ell\, \sigma^{\text{in}}_\ell\,\sigma^{\text{out}}_\ell
  \,,\\
  H_{\mathrm{boundary}}
  =
  &{\ } - \sum_{\ell \in L_\partial} h_\ell\, 
  \left(\sigma^{\text{in}}_\ell + \sigma^{\text{out}}_\ell\right)
  \,.
\end{align}\label{eq:model}%
\end{subequations}
%where $N_{\mathrm{gates}}$ denotes the total number of gates.
The function $\overline T_g$ is the negation of the function $T_g$
that encodes the truth table of gate $g$: if the input and output
spins satisfy the gate, $T_g=1$ ($\overline T_g=0$), and if they do
not, $T_g=0$ ($\overline T_g=1$). $H_{\mathrm{gates}}$ penalizes
configurations (\emph{i.e.}, costs energy $\Delta_g>0$) if the input and
output bits (or their spin equivalents) violate the truth table of
gate $g$.
%Here we take the limit $\Delta\to\infty$, working within the space of
%states where each gate itself is satisfied, but not necessarily is
%compatible with its neighbors.
The neighbor-gate compatibility is implemented through
$H_{\mathrm{links}}$, containing ferromagnetic terms of strength
$J_\ell>0$ that enforce alignment between connected inputs and outputs
of neighboring gates (sharing a link $\ell$). (Notice that we allow
for the most general form of the Hamiltonian that involves gate- and
link-dependent couplings, a choice that is relevant to the discussion
of Sec.~\ref{sec:thermodynamics} below.) Finally,
$H_{\mathrm{boundary}}$ describes local bias fields $h_\ell$ that fix
the states of a subset of twin spins on a boundary link ($\ell \in
L_\partial$) when $h_\ell\to \infty$. These biases fix the boundary
conditions of inputs or outputs appropriate for the specific
computation (multiplication or factoring) implemented by the system.

To be more precise, by fixing $S=0$, the bits of $a$ and $b$, and
ancillary inputs $c = 0$ and outputs $c'=0$ using external fields
$h_\ell$ in Eq.~\eqref{eq:model}, the model performs the
multiplication $S'=a\times b$ upon reaching the minimum energy
state. Alternatively, factorization of the $2L$-bit integer $S'$ is
implemented by appropriate selection of external fields $h_\ell$ that
fix the bits of $S'$ and $c = c' = 0$, while leaving the spins
belonging to $a$ and $b$ free. In the case of multiplication, the
ground state is uniquely determined by the boundary, since there is
only one product $S'$ for given factors $a$ and $b$. On the other
hand, depending on the number $S'$ to be factorized, there may be
multiple solutions. For a semi-prime $2L$-bit number that is the
product of two $L$-bit prime numbers, there are two solutions
(differing by where the numbers $a$ and $b$ settle along the
boundaries, since $a\times b= b\times a$).

As already mentioned in the introduction, this paper was motivated by
the astronomically slow (double exponential) time-to-solution found by
our earlier annealing studies of the the relaxation into the ground
state of the statistical mechanics model encoding the multiplication
circuit presented above. The initial goal of those studies was to use
this model to factorize semi-prime numbers.  More interesting for
applications to physical systems is that the above computationally
inspired model also displays double exponentially slow relaxation
rates for open boundary conditions ($h_\ell=0$), a result we discuss
in detail in Sec.~\ref{sec:annealing}.  Hereafter we will concentrate
on this free-boundary case, for which we prove that no thermodynamic
phase transition occurs down to zero temperature (see below).
%Taken together, the next two sections
%of this paper thus argue for a novel example of dynamical freezing
%without disorder or a thermodynamic transition.

\vspace{.25cm}

\section{Thermodynamics}\label{sec:thermodynamics}

In this Section, we derive the exact expression for the partition
function of the model defined in Sec.~\ref{sec:model} and we show that
the system lacks a finite-temperature thermodynamic phase
transition. In what follows, we work with open boundary conditions
($h_\ell=0$). The partition function (omitting one layer of
ferromagnetic bonds in the wires at the outputs of the whole circuit)
is given by
\begin{widetext}
\begin{equation}
  Z'
  =
  \sum_{\{\sigma^{\text{in}}_\ell\},\{ \sigma^{\text{out}}_\ell\}}
  \prod_{g=1}^{N_{\mathrm{gates}}}
  e^{-\beta\Delta_g {\overline T}_g[\sigma^{\text{in}}(g), \sigma^{\text{out}}(g)]}
  \;
  e^{\beta \sum_{\ell\in w^{\rm in}(g)} J_\ell\, \sigma^{\text{in}}_\ell\,\sigma^{\text{out}}_\ell}
  \;.
\end{equation}
\end{widetext}
This expression can be computed via transfer matrices, if we properly
slice the lattice in a sequence of layers. These layers contain a
single gate $g$, as we depict in Fig.~\ref{fig:layers}. An $L$-bit
multiplier can be thus sliced into $2 L^2$ layers, one for each
gate. The layer has two boundaries, which enclose the gate $g$. For
all spins not attached to links in $w^{\text{in}}(g) \cup
w^{\text{out}}(g)$, the transfer matrix acts trivially; we denote the
spins on this trivial line by $\bar\sigma(g)$. Therefore the transfer
operator can be written as ${\cal T}_g=t_g\otimes
\openone_{\bar\sigma(g)}$, where $t_g$ acts only on the spins on the
links attached to the gate. Specifically, the layer functions as a
transfer matrix between the spins $\sigma^{\text{out}}_\ell, \ell \in
w^{\text{in}}(g)$ and the spins $\sigma^{\text{out}}_{\ell'}, \ell'
\in w^{\text{out}}(g)$. Explicitly, denoting
$\{\ell_1,\ell_2,\ell_3,\ell_4,\ell_5\}=w^{\text{in}}(g)$ and
$\{\ell'_1,\ell'_2,\ell'_3,\ell'_4,\ell'_5\}=w^{\text{out}}(g)$, $t_g$
can be written as
\begin{widetext}
\begin{equation}
\label{eq:tg}
t_g
=
\sum_{\sigma^{\text{out}}_{\ell^{\ }_i},\,\sigma^{\text{out}}_{\ell'_i}=\pm 1}
|\sigma^{\text{out}}_{\ell'_1},
\sigma^{\text{out}}_{\ell'_2},
\sigma^{\text{out}}_{\ell'_3},
\sigma^{\text{out}}_{\ell'_4},
\sigma^{\text{out}}_{\ell'_5}
\rangle
\left(
\sum_{\sigma^{\text{in}}_{\ell^{\ }_i}=\pm 1}
e^{-\beta\Delta_g {\overline T}_g[\sigma^{\text{in}}(g), \sigma^{\text{out}}(g)]}
  \;
  e^{\beta \sum_{\ell\in w^{\rm in}(g)} J_\ell\,\sigma^{\text{in}}_\ell\,\sigma^{\text{out}}_\ell}
  \right)
  \;
\langle
\sigma^{\text{out}}_{\ell_1},
\sigma^{\text{out}}_{\ell_2},
\sigma^{\text{out}}_{\ell_3},
\sigma^{\text{out}}_{\ell_4},
\sigma^{\text{out}}_{\ell_5}|
  \;.
\end{equation}
\end{widetext}

The partition function for open boundary conditions can then be
expressed as
\begin{equation}
  \langle\Sigma|
  \left(\prod_{g=1}^{N_{\mathrm{gates}}}
       {\cal T}_g
       \right)
       |\Sigma\rangle
       \;,
\end{equation}
where the state 
\begin{equation}
  |\Sigma\rangle
  =
  \sum_{\{\sigma\}_{5L}} |{\{\sigma\}_{5L}}\rangle
\end{equation}
corresponds to the sum over all possible configurations on the
boundary. The state $|\Sigma\rangle$ is an eigenstate of ${\cal T}_g$
with eigenvalue $\lambda_g$ given by
\begin{align}
\lambda_g &= \sum_{\sigma^{\text{in}}_{\ell^{\ }_i},\sigma^{\text{out}}_{\ell^{\ }_i}=\pm 1}
e^{-\beta\Delta_g {\overline T}_g[\sigma^{\text{in}}(g), \sigma^{\text{out}}(g)]}
  \;
  e^{\beta \sum_{\ell\in w^{\rm in}(g)} J_\ell\,\sigma^{\text{in}}_\ell\,\sigma^{\text{out}}_\ell}
  \nonumber\\
  &=
  \left(1+31e^{-\beta\Delta_g}\right)
  \;
  \prod_{\ell\in w^{\rm in}(g)}
  \left(2\cosh \beta J_\ell\right)
  \;,
\end{align}
as follows from the non-trivial part $t_g$ of the transfer operator in
Eq.~(\ref{eq:tg}). (The factor of $31=2^5-1$ accounts for all the
configurations that pay energy for violating the truth table.) The
partition function $Z'$ is thus given by
\begin{align}
  Z'
  =\langle\Sigma|\Sigma\rangle\;\prod_{g=1}^{N_{\mathrm{gates}}} \lambda_g
  = 2^{5L}\;\prod_{g=1}^{N_{\mathrm{gates}}} \lambda_g
  \;.
\end{align}
Recall that we omitted the $5L$ ferromagnetic bonds at the outputs of
the circuit; accounting for those links yields
\begin{align}
  Z
  = Z' \; \prod_{\ell \in L^{\rm out}_\partial}\left(2\cosh \beta J_\ell\right)
  \;,
\end{align}
or equivalently, factoring out the contributions from all the gates
and the links,
%\begin{subequations}
\begin{align}
  Z
  &= 2^{5L}\; Z_{\rm gates}\; Z_{\rm links}
  \nonumber\\
  &= 2^{5L}\; 
  \left[\prod_{g=1}^{N_{\mathrm{gates}}} \left(1+31e^{-\beta\Delta_g}\right)\right]
  \;
  \left[\prod_\ell\left(2\cosh \beta J_\ell\right)\right]
  \;.
\end{align}
%\end{subequations}
The resulting free energy of the model then reads,
\begin{align}
  \label{eq:free-energy}
  \beta F
  =\;& - {5L}\ln 2
  \nonumber\\
  &-
  \left[\sum_{g=1}^{N_{\mathrm{gates}}}
    \ln\left(1+31e^{-\beta\Delta_g}\right)\right]
  \nonumber\\
  &-
  \left[\sum_\ell
    \ln\left(2\cosh \beta J_\ell\right)\right]
  \;.
\end{align}

Notice that this free energy has no singularities as a function of
$\beta$ and hence, the spin model with open boundary conditions
($h_\ell = 0$) displays {\em no} finite temperature thermodynamic
transition, {\em independent} of the values of the couplings
$\Delta_g, J_\ell$. Moreover, the free energy is independent of the
specific truth table assigned to each of the gates. In particular, the
lack of a thermodynamic transition survives for a model in which the
row of $\mathsf{RCARRY}-\mathsf{CORNER}$ gates is replaced by a row of
$\mathsf{RCARRY}-\mathsf{LCARRY}$, and the single (double) U-turns at
the top (bottom) of the multiplication circuit are removed
(\emph{i.e.}, the corresponding $J_\ell$ couplings set to zero.) While
the resulting circuit no longer implements multiplication, it is this
``multiplication-circuit inspired'' (MCI) model that can be regarded
as a translationally invariant physical system with short ranged
interactions. Below, we show that both the multiplication circuit and
the MCI model with open boundary conditions display remarkably slow
relaxation into their ground states.

\section{Thermal Annealing}\label{sec:annealing}

In this Section we use classical thermal annealing to study the
relaxation dynamics of the MCI model with open boundary conditions. We
shall focus on the case of a uniform ferromagnetic interaction
$J_\ell=J$ and $\Delta_g\to \infty$, for which all gates satisfy their
truth tables. In this case, the dynamics is carried out via a
Metropolis algorithm that flips a gate state $q$, satisfying the truth
table, into another satisfying state $q'$.

We start from a random initial spin configuration and lower the bath
temperature from $T_0$ to $T=0$ during a time $\tau$ according to:
\begin{equation}
  \label{eq:protocol}
  T(t)=T_0 \left(1-t/\tau\right)
  \;.
\end{equation}
We define $E_\tau(T)$ to be the value of the energy per link when the
time-dependent temperature $T(t)$ reaches a given value $T$. In the
limit of infinitely slow annealing, {\it i.e.}, $\tau\to\infty$, one
has $E_\infty(T)=E_{\mathrm{thermal}}(T)$. Using
Eq.~(\ref{eq:free-energy}), the total thermal energy per link is
\begin{equation}
  \label{eq:thermal_exact}
  E_{\mathrm{thermal}}(T) = J\left[1-\tanh (J/T)\right]
  \;,
\end{equation} 
where we shifted the ground state energy to 0, which is convenient for
the annealing studies below.

\begin{figure}[t]
\includegraphics[width=3.5in]{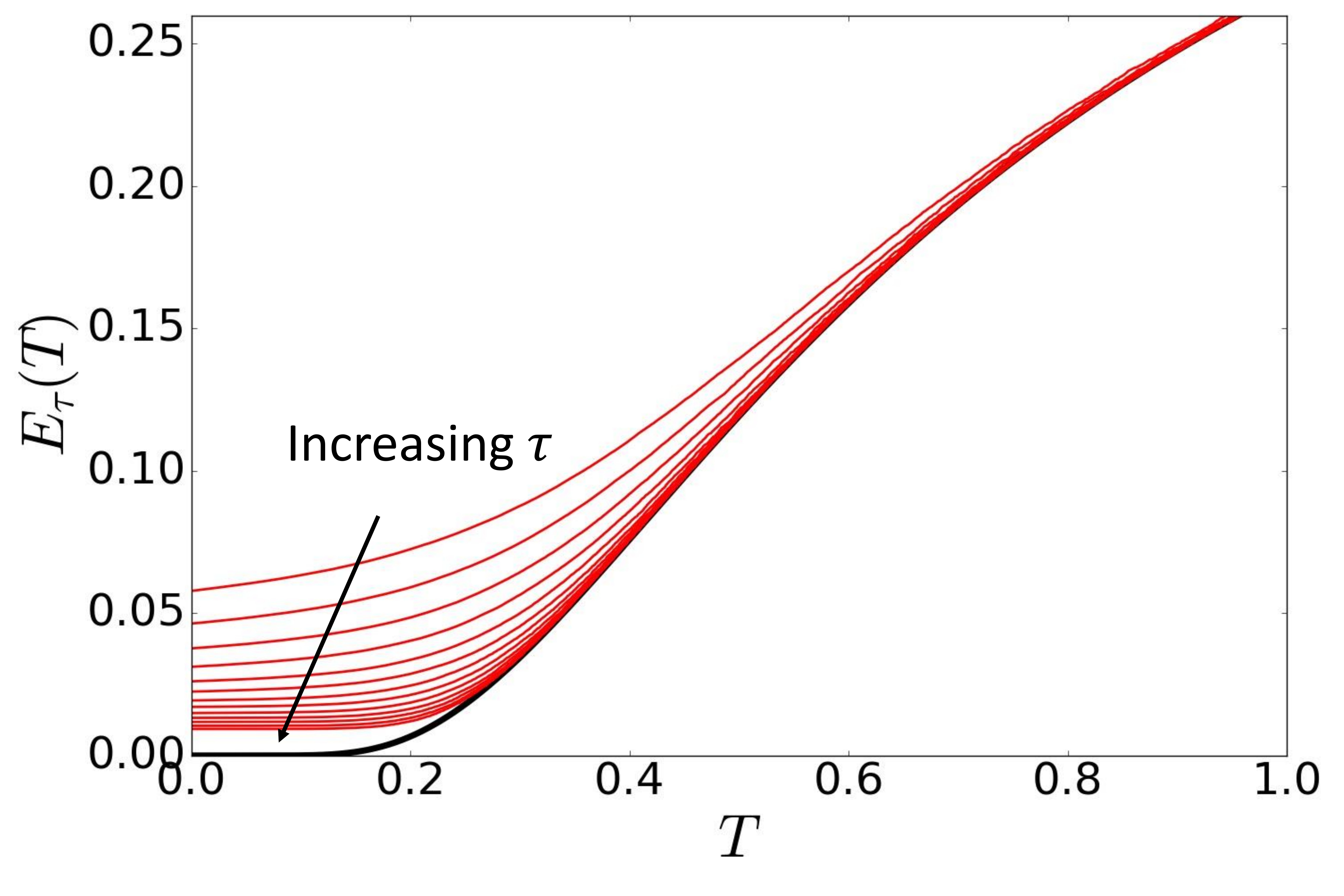}
\caption{Energies $E_\tau(T)$ for a multiplication circuit with
  $L=512$, as a function of the instantenous temperature $T$ for
  cooling times $\tau$ varying from $2^{9}$ to $2^{21}$ are shown in
  red. The simulations are carried out for $J=1/2$ and $T_0=1$. The
  thermodynamic (equilibrium) energy [Eq.~(\ref{eq:thermal_exact})] is
  shown in black.}
\label{fig:annealing_time}
\end{figure}

Figure~\ref{fig:annealing_time} shows $E_\tau(T)$ for the $L=512$
multiplier and different values of $\tau$. The solid black line
corresponds to the equilibrium result in
Eq.~\eqref{eq:thermal_exact}. Red lines show the numerical results for
$E_\tau(T)$ for values of $\tau$ in the range from $2^{9}$ to
$2^{21}$. Notice that $E_\tau(T)$ monotonically decreases, yet the
system is unable to reach its ground state for finite annealing times,
falling out of equilibrium. The minimum or residual energy $E_\tau(0)\sim
1/\ell_\tau^2$ defines a length scale $\ell_\tau$ that measures the
typical distance between defects in the frozen state.

\begin{figure*}

\begin{minipage}[t]{0.48\textwidth}
\includegraphics[width=\textwidth]{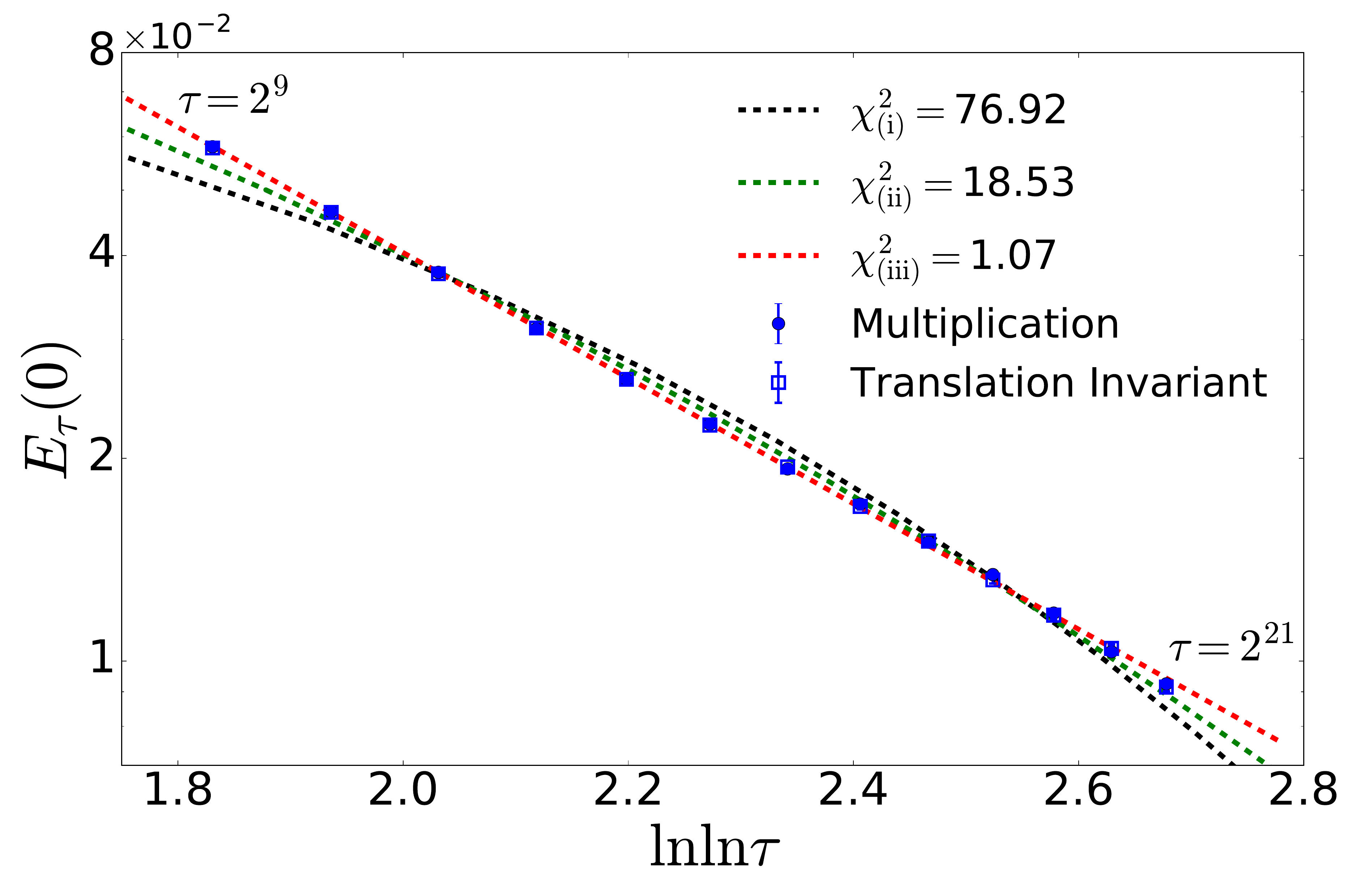}
\caption{Residual energy $E_\tau(T=0)$ at the end of the cooling
  protocol as a function of annealing time $\tau$ in the range $2^{9}$
  to $2^{21}$ for the MCI model and the multiplication circuit with $L
  = 512$ and open boundaries. Fits to three functional forms of
  $E_\tau(0)$ vs annealing time $\tau$ are shown: (i) $\ln E_\tau(0) =
  -\alpha_{\mathrm{(i)}}\ln\tau +\gamma_{\mathrm{(i)}}$ (black) with
  $\alpha_{\mathrm{(i)}} =0.21$ and $\gamma_{\mathrm{(i)}} = -1.63$;
  (ii) $\ln E_\tau(0)=-\alpha_{\mathrm{(ii)}}\sqrt{\ln\tau}
  +\gamma_{\mathrm{(ii)}}$ (green) with $\alpha_{\mathrm{(ii)}} =1.37$
  , $\gamma_{\mathrm{(ii)}} = 0.51$; and (iii) $\ln E_\tau(0) =
  -\alpha_{\mathrm{(iii)}} \ln\ln\tau+\gamma_{\mathrm{(iii)}}$ (red)
  with $\alpha_{\mathrm{(iii)}} =2.14$ , $\gamma_{\mathrm{(iii)}} =
  1.08$. The $\chi^2$ for the three fits are shown in the legend. The
  best fit is to form $\mathrm{(iii)}$, with a $\chi^2$ close to 1. We
  note that the simulation results for the multiplication circuit and
  the MCI models fall on top of one another within the error bars of
  the data.}
  \label{fig:multiply_anneal}
\end{minipage}%
\hfill%
\begin{minipage}[t]{0.48\textwidth}

  \includegraphics[width=\textwidth]{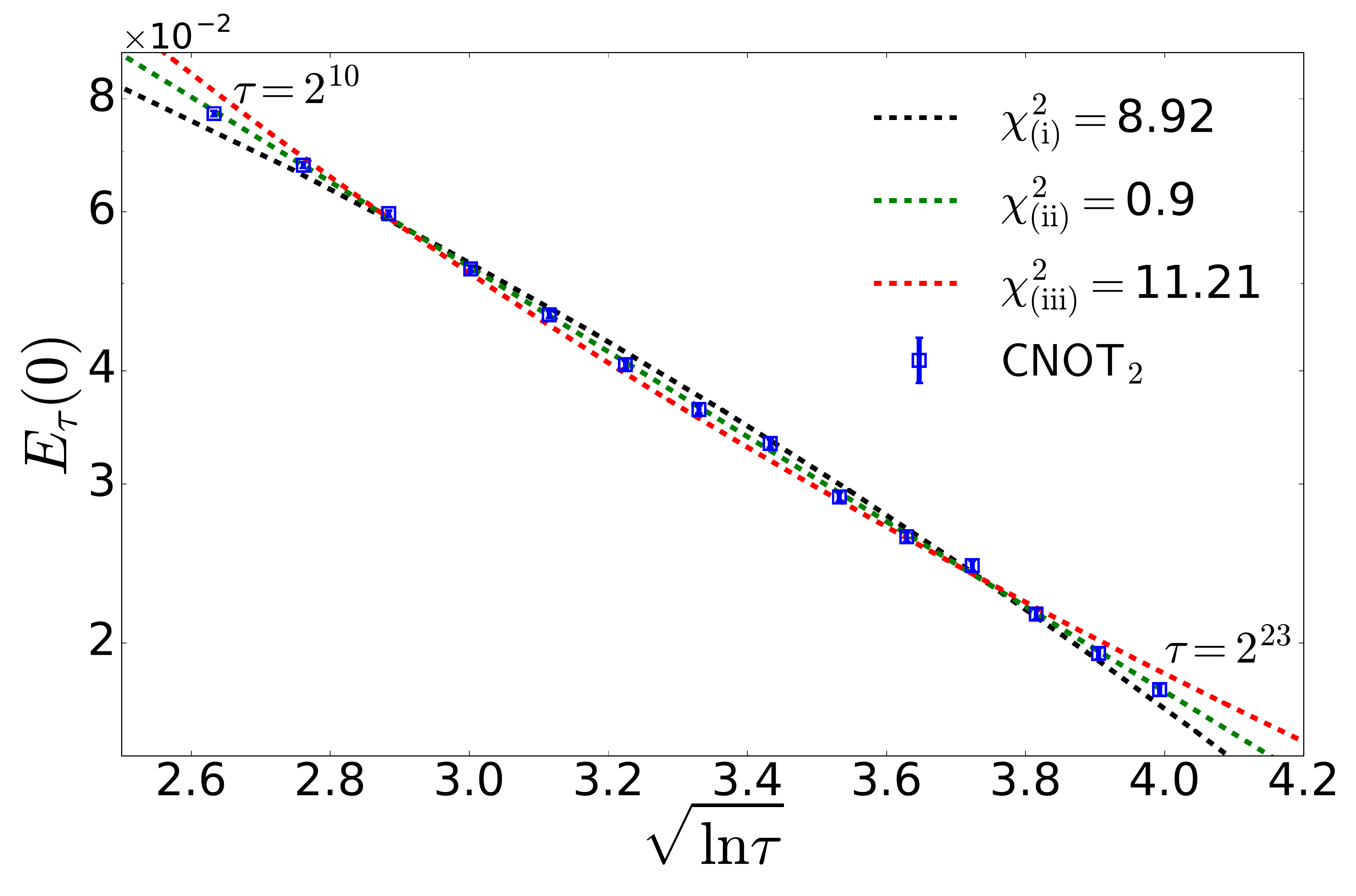}
  \caption{Residual energy $E_\tau(T=0)$ at the end of the cooling
    protocol as a function of annealing time $\tau$ in the range
    $2^{10}$ to $2^{23}$ for the $\mathsf{CNOT}_2$ circuit with $L =
    512$ and open boundaries. Fits to three functional forms of
    $E_\tau(0)$ vs annealing time $\tau$ are shown: (i) $\ln E_\tau(0)
    = -\alpha_{\mathrm{(i)}}\ln\tau +\gamma_{\mathrm{(i)}}$ (black)
    with $\alpha_{\mathrm{(i)}} =0.16$ and $\gamma_{\mathrm{(i)}} =
    -1.48$; (ii) $\ln E_\tau(0)=-\alpha_{\mathrm{(ii)}}\sqrt{\ln\tau}
    +\gamma_{\mathrm{(ii)}}$ (green) with $\alpha_{\mathrm{(ii)}}
    =1.08$ , $\gamma_{\mathrm{(ii)}} = 0.29$; and (iii) $\ln E_\tau(0)
    = -\alpha_{\mathrm{(iii)}} \ln\ln\tau+\gamma_{\mathrm{(iii)}}$
    (red) with $\alpha_{\mathrm{(iii)}} =1.77$ ,
    $\gamma_{\mathrm{(iii)}} = 0.93$. The $\chi^2$ for the three fits
    are shown in the legend. The best fit is to form $\mathrm{(ii)}$,
    with a $\chi^2$ close to 1.  }
  \label{fig:cnot_anneal}
\end{minipage}%
\end{figure*}

To analyze the data we consider fits to three functional forms of
$E_\tau(0)$ vs. annealing time, $\tau$: (i) $\ln E_\tau(0) \sim
-\ln\tau$, (ii) $\ln E_\tau(0) \sim -\sqrt{\ln\tau}$, and (iii) $\ln
E_\tau(0) \sim -\ln\ln\tau$. These arise from three models of
relaxation via activation over a barrier corresponding, respectively,
to different dependences of the barrier $\Delta_B(\ell_\tau)$ on the
length scale $\ell_\tau$: (i) $\Delta_B(\ell_\tau)\sim {\rm
  constant}$, (ii) $\Delta_B(\ell_\tau)\sim \ln\ell_\tau$, and (iii)
$\Delta_B(\ell_\tau)\sim \ell_\tau^a$, $a>0$.

In Figure~\ref{fig:multiply_anneal} we plot $E_{\tau}(0)$ as a
function of $\tau$ for both the multiplication circuit and for the MCI
model for $L = 512$. The fits to the three types of annealing behavior
along with the corresponding reduced chi-squared statistic $\chi^2$,
which we use as a figure of merit to evaluate the quality of each fit,
strongly support model (iii), namely $\ln E_{\tau} (0) \sim
-\ln{\ln\tau}$, which follows from an energy barrier that scales as
$\Delta_B(\ell_\tau)\sim \ell_\tau^a$, with $a\sim 1$.

The energy $E_\tau(0)$ reached at the end of the protocol for a given
$\tau$ can be matched to the thermodynamic value of
$E_{\mathrm{thermal}}(T)$ for some $T>0$; we define $\tau(T)$ as the
time scale needed in order that $E_\tau(0)$ reaches
$E_\tau(0)=E_{\mathrm{thermal}}(T)$. In that case, $\ell_\tau=\xi(T)$,
the thermal correlation length, which, given that the model displays
no finite temperature transition, only diverges at zero temperature as
$\xi\sim e^{-J/T}$.  This translates into a relaxation time for
equilibration of the double exponential form:
\begin{equation}
  \tau \sim
  \exp\left[\exp(J/T)\right].
  \label{eq:main-result}
\end{equation}

The remarkable appearance of such a slow dynamical relaxation time in
a system for which we prove that there is no finite-temperature
thermodynamic phase transition is the main result of our work. While
we do not have a detailed analytical understanding of the origin of
this behavior, in Sec.~\ref{sec:defects} we will argue that its origin
can be traced back to the amplification of errors induced downstream
of a single-bit defect in the course of the computation.

\begin{figure}[t]
\includegraphics[width=3.5in]{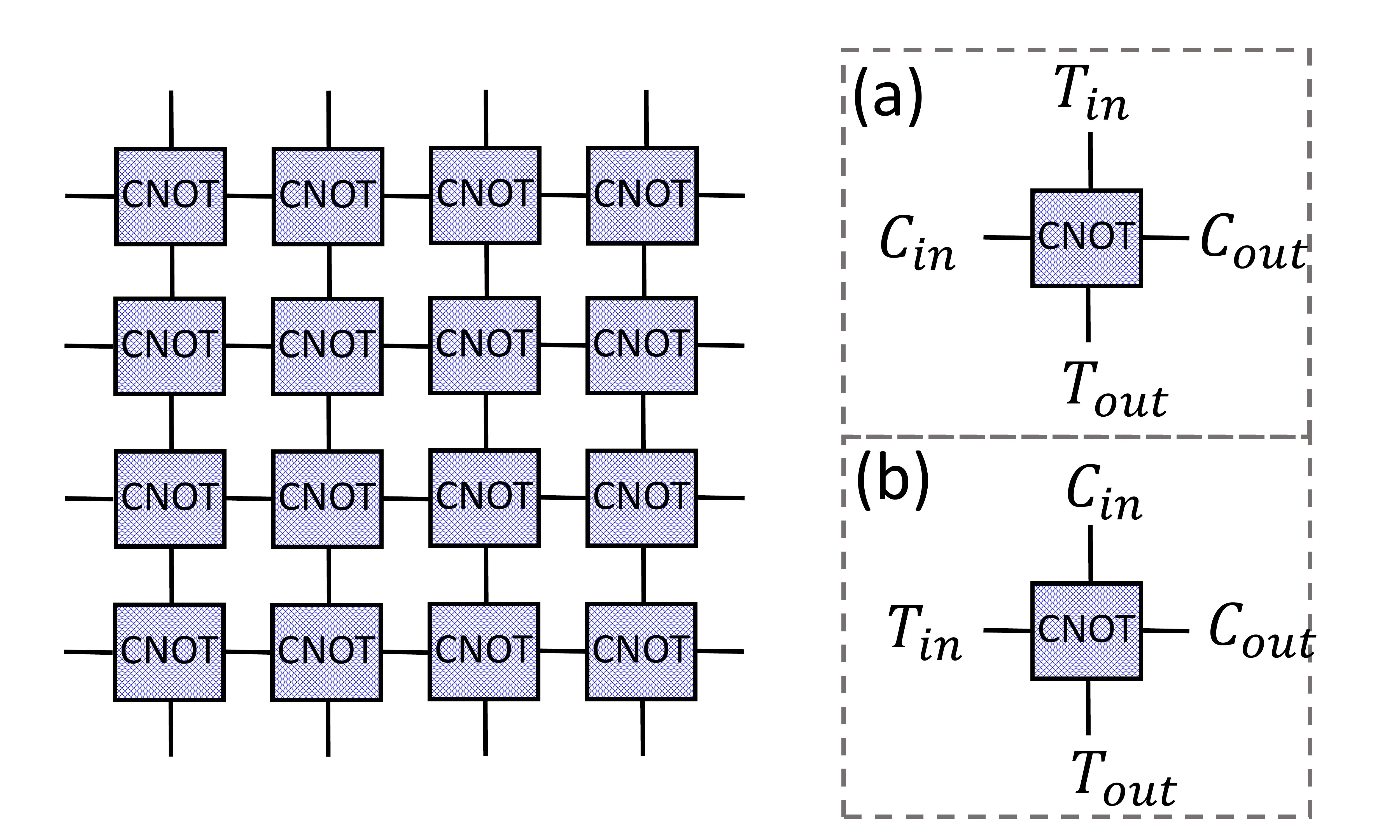}
\caption{Left panel: Translationally invariant spin systems
  constructed by $\mathsf{CNOT}$ gates. Right panel: We show two
  different ways of wiring $\mathsf{CNOT}$ gates in a square lattice
  array. (a) In the first model, referred to as $\mathsf{CNOT}_1$, the
  output control ($C$) and output target ($T$) bits of a given gate
  are connected to the input control and the input target bits of a
  neighboring gate, respectively. (b) In the second $\mathsf{CNOT}_2$
  model, the wires are switched: the output control ($C$) and output
  target ($T$) bits of a given gate are connected to the input target
  and input control bits of a neighboring gate, respectively.
}\label{fig:cnotcircuit}
\end{figure}

For comparison, and in order to sharpen the discussion of
Sec. \ref{sec:defects}, we shall consider two additional
translationally invariant spin systems derived from computational
models that display the behaviors of cases (i) and (ii) above. These
simpler computational models, shown in Fig.~\ref{fig:cnotcircuit},
represent two different ways of wiring $\mathsf{CNOT}$ gates in a
square lattice array. The first model, referred to as
$\mathsf{CNOT}_1$, is illustrated in Fig.~\ref{fig:cnotcircuit}(a). In
$\mathsf{CNOT}_1$ the output control ($C$) and output target ($T$)
bits of a given gate are connected to the input control and the input
target bits of a neighboring gate, respectively. In the second
$\mathsf{CNOT}_2$ model, shown in Fig.~\ref{fig:cnotcircuit}(b), the
wires are switched: the output control ($C$) and output target ($T$)
bits of a given gate are connected to the input target and input
control bits of a neighboring gate, respectively. While, as in the
case of the MCI model, one can prove that neither of the resulting
statistical mechanics models displays a finite temperature
thermodynamic transition, the simple change in the wiring drastically
changes the relaxational dynamics of the corresponding models. In
particular, the freezing energy $E_\tau(0)$ vs. annealing time behaves
as cases (i) and (ii) for $\mathsf{CNOT}_1$ and $\mathsf{CNOT}_2$,
namely, $\ln E_\tau(0) \sim -\ln\tau$, and $\ln E_\tau(0) \sim
-\sqrt{\ln\tau}$, respectively.

Indeed, Figure~\ref{fig:cnot_anneal} shows $E_{\tau}(0)$ as a function
of $\tau$ for a lattice of size $L = 512$ for the $\mathsf{CNOT}_2$
model. As in the case of the multiplication and MCI models, we show
the three corresponding fits to the three kinds of annealing
behaviors, along with the $\chi^2$ for each of these fits. As
advertised, the best fit to the data corresponds to case (ii), namely
$\ln E_{\tau} (0) \sim -\sqrt{\ln\tau}$.

\section{Defects and propagation of errors}
\label{sec:defects}

The three classes of dynamics encountered above can be rationalized by
considering the nature of the defects --- links where input and output
spins are mismatched --- and the energy barriers associated with the
process of healing them. Even though a single defect only costs energy
$2J$, it affects an extended region of downstream bits as the
computation proceeds. Since the truth tables of all gates are always
satisfied, another way of visualizing errors generated by a single
defect is to follow the changes induced by the computation in the
state, $q=0,1,...{2^5}-1$, of each of the gates from those in a ground
state in the absence of defects (the "parent state").

\begin{figure*}[t]
\includegraphics[width=7in]{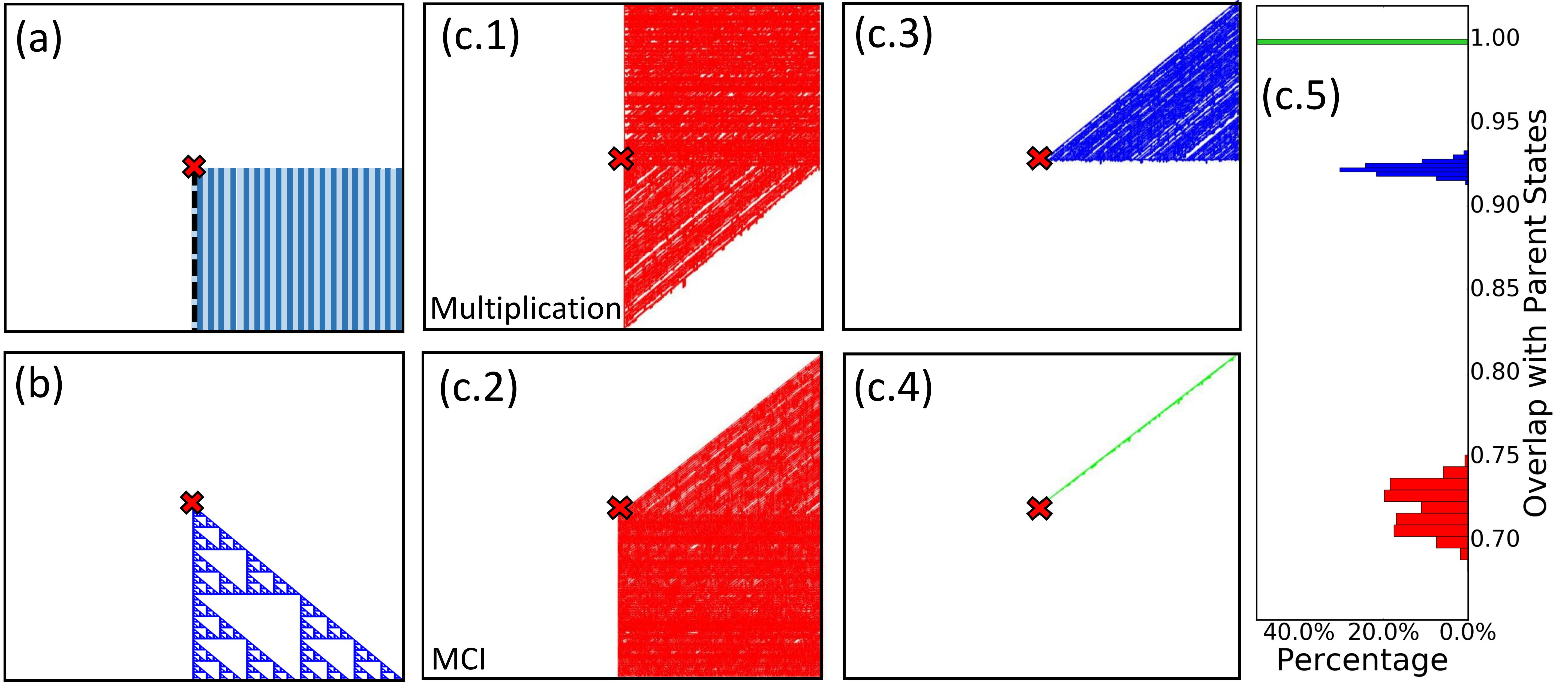}
\caption{Downstream propagation of errors due to a single defective
  link placed at the center of a system of size $L = 512$ for the four
  models: $\mathsf{CNOT}_1$, $\mathsf{CNOT}_2$, multiplication
  circuit, and MCI. Colored areas indicate the regions of errors in
  which the states of gates differ from those in the reference ground
  state (defect free state). (a) Error propagation in the
  $\mathsf{CNOT}_1$ model. The errors initiated in a spin-flip at the
  control $C$-input form independent columns, which can be healed
  sequentially as detailed in the text, whereas spin-flips of the
  target $T$-input generate the vertical dashed black line. (b) Error
  propagation in the $\mathsf{CNOT}_2$ model. The error region is
  patterned in a Sierpinski gasket independent of whether the error is
  initiated by a flip in the control or target input spins. (c) Error
  propagation in the multiplication circuit and in MCI models, both of
  which feature different shaped regions of errors depending on which
  of the five input spins of the $\mathsf{LCARRY}$ gate at the center
  of the lattice initiate the error: the red regions of equal area in
  panels (c.1) and (c.2) originate from a defect in the $a$-input of
  the gate for the multiplication and MCI models, respectively; the
  blue region in panel (c.3), which is virtually the same for the two
  models, originates from defects in either $b$- or $c$-input spins;
  and the green line in panel (c.4) is featured in both models and
  originates from a defect in the $S$-input spin of the
  $\mathsf{LCARRY}$ gate. The histogram in panel (c.5) shows the
  distributions of the three types of defects in the multiplication
  and MCI models (color-coded as in panels (c.1)-(c.4)), computed for
  1024 randomly chosen parent states. In panel c.5, the green line
  representing the overlap with the error induced by a defect in the
  $S$-input to the gate should go all the way to 100$\%$ but was cut
  off in order to set a reasonable scale for the rest of the
  figure. (We note that defects in the simpler $\mathsf{CNOT}_1$ and
  $\mathsf{CNOT}_2$ models do not depend on the parent state.)}
\label{fig:Error_Propagation}
\end{figure*}

More precisely, we start with a zero-energy (\emph{i.e.}, ground)
state and create an excited state with minimum energy $2J$ by flipping
one of the input spins of a gate at the center of the system and
realigning the rest of the spins downstream of the error, so that all
bonds but the original defective one are satisfied. To visualize the
buildup of errors in this excited state, we compute its overlap with
the ``parent'' ground state. Figure~\ref{fig:Error_Propagation}
exemplifies single defects and their downstream bit flips induced by
the computation for the four circuits $\mathsf{CNOT}_1$,
$\mathsf{CNOT}_2$, multiplication, and MCI. The single error leads to
the presence of two distinct regions $A$ (white region in
Fig.~\ref{fig:Error_Propagation}) and $B$ (colored region in
Fig.~\ref{fig:Error_Propagation}) that, while containing no error
themselves, correspond to different ground states. We remark that
there is no interfacial energy penalty --- the cost is again only the
$2J$ paid at the defective link. To relax the defect and reach a
global ground state one must flip the entire region $A$ or $B$.

In Fig.~\ref{fig:Error_Propagation}(a) we consider the case of the
$\mathsf{CNOT}_1$ circuit. We note that when the defect is induced in
the input target ($T$) spin (bit) the line of downstream defects
(dashed black line in panel (c.1)) can be healed by moving the origin
of the defect line downward to the lower boundary of the circuit with
no energy cost, as in the case of a domain wall in a one-dimensional
Ising model. When defects are initiated from a flip in the input
control ($C$) spin the region $B$ is built out of disconnected single
lines of flipped bits, which can be healed sequentially. In this case,
the defect can be moved vertically towards the boundary by generating
one additional defect at cost $2J$. The additional defect can then be
moved horizontally all the way to the boundary at no extra
cost. Therefore, the defects can effectively propagate vertically by
paying a cost $2J$. This is therefore an example of class (i), with a
constant energy barrier $2J$.

In Fig.~\ref{fig:Error_Propagation}(b) we show the corresponding
downstream effect of a single defect initiated at the $C$-input in the
$\mathsf{CNOT}_2$ circuit for which region $B$ is a Sierpinski gasket
(a fractal), identical to that found in the triangular plaquette model
of Ref.~\onlinecite{Newman-Moore1999}. (The same pattern also occurs
when the errors are initiated from a defect in the $T$-input.) We
remark that the $\mathsf{CNOT}_2$ circuit provides a new realization
of this type of fractal structure on a square, instead of triangular,
lattice. As in the triangular plaquette
model~\cite{Newman-Moore1999,Garrahan2000}, defects can be healed in a
hierarchical manner by overcoming energy barriers that are logarithmic
in the linear size of the triangular region, thus providing an example
of class (ii). This hierarchical relaxation leads to a qualitatively
slower super-Arrhenius characteristic relaxation time as function of
temperature of the form $\tau\sim \exp{1/T^2}$.

Finally, the downstream effects of a single defect in the
multiplication circuit and MCI model (at the input of a
$\mathsf{LCARRY}$ gate at the center of the system) are shown in
Fig.~\ref{fig:Error_Propagation}(c). In this case there are three
kinds of defects, depending on which of the inputs ($a$; $b$ or $c$;
or $S$) to a gate is flipped to initiate the error.
Fig.~\ref{fig:Error_Propagation}(c) shows the different shapes of the
regions that are misaligned with the parent state for the different
types of defects. For all but $S$ defects, errors spread into a
two-dimensional wedge. As detailed in the figure, the multiplication
circuit and MCI model show the same behavior when the errors are
initiated from the $b$-, $c$- or $S$-inputs, but display error regions
of the same size but of different orientation for defects originated
from the $a$-input. We note that, unlike the $\mathsf{CNOT}$ models,
here the overlaps do depend explicitly on the parent state. Panel
(c.5) of Fig.~\ref{fig:Error_Propagation}(c) shows the distribution of
overlaps for each type of defect computed for 1024 randomly chosen
parent states.

\begin{figure}[t]
\includegraphics[width=3.5in]{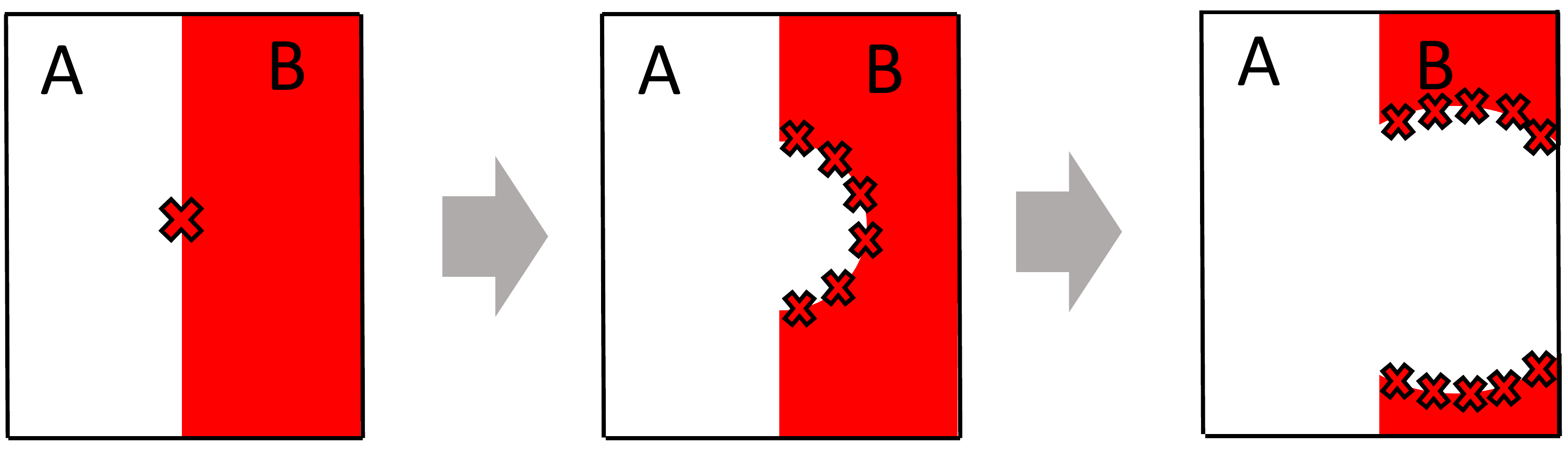}
\caption{Panels from left to right: cartoon of the defect relaxation
  process, illustrating an energy barrier $\propto \xi$ required to
  flip the right side of the lattice to state $A$ and evaporate the
  defect.  }\label{fig:cartoon}
\end{figure}

Qualitatively we expect that healing for the case in
Fig.~\ref{fig:Error_Propagation}(c) should proceed as depicted in the
cartoon in Fig.~\ref{fig:cartoon}. In order to flip region $B$, one
progressively increases the bulge of region $A$ invading $B$. The
boundary or interface of the bulge must contain additional defects and
hence the barrier scales as the length scale of the bulge, where
defects are created to produce the curvature. It is this scaling of
the energy barrier with the length of the buldge that makes this model
qualitatively different from classes (i) and (ii), and consistent with
the scaling of the barrier within class (iii). This cartoon
rationalizes the surprisingly slow relaxation times observed in the
data presented in Fig.~\ref{fig:multiply_anneal}, which are consistent
with the double exponential behavior of the rate with the inverse
temperature in Eq.~(\ref{eq:main-result}). We summarize the results of
this Section in Table~\ref{tab:scaling}.

\begin{table}
\centering
\begin{tabular}{c |c c c }
\hline
Type & (i)&(ii)&(iii) \\
\hline
Circuit & \;\;\;\;$\mathsf{CNOT}_1$ \;\;\;\;& $\mathsf{CNOT}_2$& multiplication/MCI\\
$\Delta_B$  & $\mathcal{O}(1)$ & $\mathrm{ln}\xi$ & $\xi$\\
$d_f$ & 1 & $1.5849$& $\sim 2$\\
 $\tau$   &
 $ \mathrm{exp}(1/T)$ & $\mathrm{exp}(1/T^2) $& $ \mathrm{exp}(\mathrm{exp}(1/T))$ \\
\hline
\end{tabular}
\caption{Summary of the three types of relaxation rates encountered in
  this work. For each of these relaxation types we show: an example of
  a circuit displaying that type of behavior; the scaling of the
  barrier height $\Delta_B$ with correlation length $\xi$; the fractal
  dimension $d_f$ of the region of errors due to a single defect; and
  the resulting temperature dependence of the relaxation rate $\tau$.}
\label{tab:scaling}
\end{table}

\section{Conclusion}\label{sec:conclusion}

We have introduced a translationally invariant two-dimensional spin
model, inspired by a reversible multiplication circuit, and studied
its thermodynamic and dynamic properties. We have proven that this
model displays no phase transition at any finite temperature.  Even
though the thermodynamic behavior is trivial, healing thermally
excited defects is extremely difficult. Single defects cost little
energy but healing them requires flipping an extensive number of spins
(bits).  In turn this leads to ultraslow dynamics with a remarkably
long relaxation time that scales as a {\em double exponential} of the
inverse temperature.

%It is also important to stress a point already made in the introduction, namely that, contrary to naive intuition, the double exponential behavior of the relaxation time is independent of boundary conditions. 
It is important to stress that, since for different boundary
conditions the same statistical mechanics model represents
computations of different complexity --- for example, multiplication
vs. factoring --- an efficient algorithm for reaching the ground
states and thus solutions to these problems should generically lead to
qualitatively different times-to-solution. However, our results show
that thermal annealing leads to astronomically long times-to-solution,
\emph{independent of boundary conditions, and thus independent of the
  complexity of the computation}. Thermal annealing is therefore
ineffective in solving even P-class computational problems, such as
multiplication. On the other hand, thermal annealing is the generic
path to relaxation for physical systems in contact with a thermal
reservoir. From this point of view, some computation-inspired
statistical mechanics models may provide a framework for identifying
disorder-free systems with ultra-slow dynamics despite of the absence
of a finite temperature phase transition.

\vspace{.25cm}

\begin{acknowledgments}
  We thank L.~Cugliandolo, A.~de~Souza, and O.~Pfeffer for useful
  discussions. S.~K. was partially supported through the Boston
  University Center for Non-Equilibrium Systems and Computation
  (A.~E.~R). This work was supported by DOE Grant No. DE-
  FG02-06ER46316 (C.~C.). Numerical calculations were performed on the
  Boston University Shared Computing Cluster, which is administered by
  Boston University Research Computing Services.
\end{acknowledgments}

%\bibliographystyle{revtex4-1}
%\bibliography{biblio}


\end{document}